\documentclass[%
 reprint,
 amsmath,amssymb,
 aps,
]{revtex4-2}

\usepackage{graphicx}
\usepackage{dcolumn}
\usepackage{bm}
\usepackage[colorlinks,citecolor=blue,urlcolor=blue,linkcolor=blue]{hyperref}
\usepackage{microtype}
\usepackage[mathlines]{lineno}
\usepackage{xcolor}
\usepackage{csvsimple} 
\usepackage{booktabs}    
\usepackage{float}

\begin{document}

\preprint{APS/123-QED}

\title{Dynamical Dark Energy from a Massive Vector Field in Generalized Proca Theory}

\author{Abhi Savaliya$^{1}$}
\email{i21ph040@phy.svnit.ac.in}
\affiliation{$^{1}$Department of Physics, Sardar Vallabhbhai National Institute of Technology, Surat 395007, Gujarat, India}

\date{\today}

\begin{abstract}

In this paper, we emphasise the recent observational findings from the Dark Energy Spectroscopic Instrument Data Release 2 (DESI DR2), which provide compelling evidence for a possible deviation from the standard $\Lambda$CDM (Cold Dark Matter) cosmology, suggesting the presence of a dynamically evolving effective dark energy component. Motivated by this, we construct a theoretical framework in which a massive cosmological vector field, $B^{\mu}$, couples non-minimally to the background curvature through marginal interactions, offering a controlled mechanism to realise the deviation from the $\Lambda$CDM model. A detailed analysis of the effective Equation of State (EoS) parameter $w(\tilde H)$ reveals a narrow region of parameter space consistent with current cosmological observations presented by DESI. The analysis yields a stringent upper bound for the coupling constant $\lambda$ to be $\lambda<2.98\times10^{-11}$, a very strong bound on mass $3.1356\times10^{-66}~\text{g} \leq m \leq 3.3627\times10^{-66}~\text{g},$  and the admissible range $-0.405 \leq \log_{10}\tilde\gamma \leq -0.38$ for which present-day value $w_0 = w(\tilde H = 1)$ corresponding to a deviation $\delta = w_0 + 1$ that lies within the region $0.107 \leq \delta \leq 0.217$. This interval reproduces the deviation inferred from the combined DESI, Cosmic Microwave Background (CMB), and Pantheon+ data, reflecting a controlled departure from the $\Lambda$CDM behaviour. In summary, the results suggest that the proposed framework of a massive vector field can account for the departure from $\Lambda$CDM behaviour highlighted by DESI in the current cosmic acceleration. Furthermore, the framework approaches the $\Lambda$CDM behaviour in late-time $t\gtrsim28$ Gyr, establishing a direct phenomenological link between the underlying parameters and the observed dynamical nature of dark energy.

\end{abstract}

\maketitle


\section{Introduction}\label{section1}
The current observations on the accelerated expansion of the Universe are commonly attributed to an anti-gravity phenomenon, termed as dark energy, most simply modelled by $``{\Lambda}"-$ \textit{cosmological constant} in the Einstein-Hilbert action, introduced as a fixed parameter rather than the emerging dynamic quantity. Furthermore, another problem tied with $\Lambda$ is the so-called \textit{old cosmological constant problem}, namely, the extreme discrepancy between the observed value of $\Lambda$ and theoretical vacuum energy expectation values from quantum field theory \cite{RevModPhys.61.1}. This profound theoretical tension has prompted the development of alternative frameworks, explored first within classical field theory, in which the observed acceleration originates from additional dynamical degrees of freedom rather than a fixed cosmological constant. Several classical field-theoretic prescriptions, such as scalar-tensor \cite{Clifton_2012}, vector-tensor \cite{B_hmer_2007}, and scalar-vector-tensor theories \cite{Heisenberg_2018}, among others, extend Einstein's General Theory of Relativity (GR) by incorporating additional dynamical fields that non-minimally couple with the background curvature. Such extensions aim to provide a dynamical origin of the observed cosmic acceleration, without fundamentally invoking the cosmological constant by hand. The most general scalar-tensor theory with second-order field equations, proposed by Horndeski \cite{Horndeski:1974wa}, provides a controlled framework for modelling dark energy dynamics, in such a manner that the theory avoids Ostrogradsky instabilities \cite{woodard2015theoremostrogradsky} while still being more general than the Brans-Dicke and quintessence theoretical frameworks. The stringent constraints on the speed of propagation of gravitational waves from GW$170817$ neutron star merger \cite{LIGOScientific:2017zic, Moore_2001}, require fine-tuning of coupling parameters, as the large landscape of the Horndeski theory is drastically reduced to a small viable landscape after imposing $c_g = 1$, unless the coupling parameters are fine-tuned \cite{KaseTsujikawa2018, Ezquiaga_2017}. In addition, the surviving models typically resemble quintessence, kinetic-essence, and kinetic braiding type frameworks \cite{KaseTsujikawa2018}. A natural generalisation of scalar-tensor theories is given by vector-tensor theories\cite{hellings1973vector, Jim_nez_2008, Jim_nez_2009}
where the cosmic acceleration is sourced by a time-like vector field based on the cosmological ansatz. All the Ostrogradsky stable vector-tensor theories can be embedded in a broader framework known as the \textit{Generalised Proca $(GP)$ theory} \cite{heisenberg2017generalisedprocatheories}. The construction of GP theory parallels that of the Horndeski theory. This correspondence manifests if one enforces/restricts $B_\mu = \nabla_\mu \phi$, for $B_\mu$ being the  $GP$ vector field and $\phi$ is the Horndeski scalar field. In this limit, the longitudinal mode of the vector field from the GP theory reproduces a subset of Horndeski interactions, highlighting the structural analogy between the two theories. In particular, GP theory generically yields $c_g\not=1$, just like Horndeski's theory, necessitating the fine-tuning of the necessary parameters. Imposing $c_g=1$ to avoid fine-tuning reduces the landscape for the GP theory; at the same time, we lose the rich phenomenology that the full theory has to offer. The late-time cosmic acceleration is studied for the concrete model $\psi^p\propto1/H$ within the full landscape $[\mathcal{L}_{2......6}]$ of GP theory in Ref. \cite{De_Felice_2017}. The observational constraints, such as data from the Cosmic Microwave Background (CMB), Baryon Acoustic Oscillations (BAOs), and Supernova Ia (SN Ia), along with the no-ghost stability of the cosmological perturbation, suggest that the late-time cosmic acceleration is favored by the equations of state $w=-1-s$ with $s=0.16_{-0.08}^{+0.08}$ over the $\Lambda CDM$ EoS $w=-1$. However, if we demand to study a small section of the entire Landscape of GP theory that is focused on the marginal couplings, we encounter a mathematical problem in the model $\psi^p\propto1/H$.

\bigskip
 
If we try to keep the coupling constants marginal in the theory, i.e., $\psi^p\propto1/H$ by setting
    \begin{align}
        &G_2= b_2X^{p_2},\\\notag
        &G_3=b_3X^{p_3},\\\notag
        &G_4 = \frac{M_p^2}{2}+b_4X^{p_4},\\\notag
        &G_5 = b_5X^{p_5}. \\\notag
    \end{align}
That is $[X]=2$, $[b_2]=[b_3]=[b_4]=[b_5] =0 $ for $p_2=p_3=p_4=1$, and $p_5=0$ yields three different values of $p\in [-1/3,0,1]$ simultaneously. Hence, for the marginal couplings, the ansatz $\psi^p\propto1/H$ violates the argument: (\textit{the energy density of $\psi$ starts to dominate over the background matter densities at the late cosmological epoch, i.e., the amplitude of the field $\psi$ should grow with the decrease of $H$} \cite{Felice_2016, De_Felice_2017}) on which it was built, for $p=-1/3,0$ and it works perfectly fine for $p=1$ but the couplings $b_4,b_5$ pickup the non-zero mass dimension, which is the vector Galileon, additionally for $p_2=1$ \cite{Tasinato_2014, Heisenberg_2014}. To address this issue, we proceed with solving the vector field equations without assuming them apriori.

\bigskip

In this work, we aim to study the deviation in the EoS parameter from the standard $\Lambda$CDM model for the current times as well as late times using the marginal couplings bounded by the observational and theoretical constraints in order to explain the deviation inferred from the combined DESI, CMB, and Pantheon+ data. Furthermore, we aim to validate the fact that enforcement of the spatial homogeneity and isotropy on the cosmological vector field makes the time-like component purely time dependent $\psi(t)$  and non-dynamical \cite{Bohnenblust:2024mou} because of which the field tensor $F_{\mu\nu}$ vanishes. But the cosmic acceleration is sensitive to the strength of the derivative self-interaction terms in the Lagrangian. We investigate all the points mentioned above in a controlled theoretical manner. In particular, realising that the Lagrangian of any theory can be written intuitively up to infinite terms by using permutations and combinations of fields and their derivatives, to construct the classical Lagrangian in a controlled way, we focus on the terms with marginal couplings. We further control our Lagrangian by requiring it to be ghost-free and stable. By performing that, it naturally imposes its embedding into $\mathcal{L}_{2,3,4,5}$ sectors of the GP theory, which is free from Ostrogradsky instabilities by construction. 

\section{The Embedding}\label{section2}

The effective Lagrangian with the dynamical field coupled non-minimally to the Ricci scalar is written as 
\begin{align}
\mathcal{L} &= R\!\left[\frac{M_p^2}{2}+\frac{1}{2}\lambda B^\mu B_\mu+\alpha\nabla_\mu B^\mu\right] 
+ \xi R_{\mu\nu}B^\mu B^\nu \notag \\
&\quad +\epsilon R_{\mu\nu}\nabla^\mu B^\nu+ \frac{1}{2}m^2B^\mu B_\mu
- \frac{1}{4}F_{\mu\nu}F^{\mu\nu}. \label{2}
\end{align}
With $F^{\mu\nu} = \nabla^\mu B^\nu-\nabla^\nu B^\mu$. The dimensional analysis shows that in $4$-dimensional spacetime with natural units ($c=\hbar=1$), we have
\[
[\mathcal{L}] = 4, \quad [d^4x \sqrt{-g}] = -4.
\]
The basic field/operator mass dimensions are
\[
[B_\mu] = 1, \quad [\nabla_\mu] = 1, \quad [F_{\mu\nu}] = 2, \quad [R] = 2, \quad [R_{\mu\nu}] = 2,
\]
and the couplings are then canonically dimensionless:
\[
[\lambda] = [\alpha] = [\xi] = [\epsilon]= 0.
\]

Further interaction couplings constructed from the vector field, its covariant derivatives, the Ricci scalar, or the Ricci tensor give rise to higher-dimensional operators. Therefore, they must be accompanied by coupling constants carrying nonzero mass dimensions, which is clearly not in alignment with the motivation of this work. The theory governed by the Lagrangian Eq. \eqref{2} shows clearly the presence of Ostrogradsky ghosts from the coupling of the Ricci scalar with the divergence of the vector field, as the term $\alpha R \nabla_\mu B^\mu$ is effectively $B^\mu\nabla_\mu R$ after integration by parts. The vector field equations will then involve a third-order derivative of the underlying metric, which is undesirable for the theory \cite{Woodard_2007,Motohashi_2015}. To ensure the absence of  Ostrogradsky instabilities by construction, we must embed the Lagrangian in $\mathcal{L}_2$, $\mathcal{L}_3$, and $\mathcal{L}_4$ and $\mathcal{L}_5$ sectors of the GP theory \cite{heisenberg2017generalisedprocatheories}, given by the set of Lagrangians, i.e., as a sum of individual sectors, 
\begin{align}
\mathcal{L} 
&= \mathcal{L}_2 + \mathcal{L}_3 + \mathcal{L}_4 +\mathcal{L}_5,
\end{align}
with,
\begin{align}
\mathcal{L}_2 &= G_2(X,\,F), \\[6pt]
\mathcal{L}_3 &= G_3(X)\,\nabla_\mu B^\mu, \\[6pt]
\mathcal{L}_4 &= G_4(X)\,R \;+\; G_{4,X}(X)\,
\Big[
(\nabla_\mu B^\mu)^2 - \nabla_\mu B_\nu \nabla^\nu B^\mu
\Big],\label{6}\\
\mathcal{L}_5 &= G_5(X) \, G_{\mu\nu} \, \nabla^\mu B^\nu \notag\\\notag
&\quad - \frac{1}{6} G_{5,X}(X) \Big[
    (\nabla_\mu B^\mu)^3 \\\notag
&\qquad- 3 (\nabla_\mu B^\mu) (\nabla_\rho B_\sigma \nabla^\sigma B^\rho) \\
&\qquad + 2 (\nabla_\rho B_\sigma \nabla^\gamma B^\rho \nabla^\sigma B_\gamma)
\Big].
\end{align}
Here, \begin{align}
X &\equiv -\frac{1}{2} B_\mu B^\mu, \\
F &\equiv -\frac{1}{4} F_{\mu\nu}F^{\mu\nu}, \\
F_{\mu\nu} &\equiv \nabla_\mu B_\nu - \nabla_\nu B_\mu.
\end{align}
We now identify our theory, see Eq.\eqref{2} as a specific realization within the generalized framework by choosing the coefficient functions $G_i$ in the following manner
\begin{align}
G_2(X,F) &= -\tfrac{1}{4}F_{\mu\nu}F^{\mu\nu} \;+\; \tfrac{1}{2} m^2 B_\mu B^\mu,\\[6pt]
G_3(X) &= \gamma\,X,\\[6pt]
G_4(X) &= \frac{M_p^2}{2} - \lambda\,X,\\[6pt]
G_{4,X}(X) &= \frac{dG_4}{dX} \;=\; -\lambda,\\
G_{5} &= \epsilon=-\alpha.
\end{align}
To establish the correspondence with Eq.\eqref{2}, note that the term inside the square bracket in Eq.\eqref{6} can be written as,
\begin{equation}
    [(\nabla_\mu B^\mu)^2 - \nabla_\mu B_\nu \nabla^\nu B^\mu] = R_{\mu\nu}B^\mu B^\nu-\nabla_\mu A^\mu.
\end{equation}
With,
\begin{equation}
    A^\mu = B^\nu\nabla_{\nu} B^\mu - B^\mu\nabla_\nu B^\nu.
 \end{equation}
 From our theory, we can safely identify the coefficient of the term $\xi R_{\mu\nu}B^\mu B^\nu$ with $G_{4, X}$.
 \begin{equation}
     -\lambda = \xi.
 \end{equation}
With all the above identification, the Ostrogradsky safe embedded Lagrangian takes the form;
\begin{align}\label{19}
\mathcal{L} &=
-\tfrac{1}{4} F_{\mu\nu}F^{\mu\nu}
+ \tfrac{1}{2} m^2 B_\mu B^\mu \notag \\[4pt]
&\quad -  \tfrac{\gamma}{2}  B_\nu B^\nu \, \nabla_\mu B^\mu \notag \\[4pt]
&\quad + \Big(\tfrac{M_p^2}{2} + \frac{\lambda}{2}B_\mu B^\mu\Big) R \notag \\\notag
&\quad - {\lambda} \Big[ R_{\mu\nu}B^\mu B^\nu-\nabla_\mu A^\mu]\\
&\quad - \alpha G_{\mu\nu} \nabla^\mu B^\nu.
\end{align}
It is important to note that any higher sectors like $\mathcal{L}_5 $ with $G_{5, X}\not=0$ and $\mathcal{L}_6$, naturally involve couplings possessing non-zero mass dimensions. Furthermore, the terms $G_{\mu\nu} \nabla^\mu B^\nu$ and $\nabla_\mu A^\mu$ do not contribute to the bulk dynamics after integration by parts, but they hold their own importance as discussed in the last section. In this work, we focus primarily on the bulk dynamics. The corresponding field equations for the Ostrogradsky stable theory (see Eq. \eqref{19}) are therefore obtained by varying the action with respect to the fields, yielding:

\bigskip

\noindent{\textbf{Vector Field equations}}
\begin{align}\label{20}
0 &= \nabla_\nu F^{\nu\mu} 
    + m^2 B^\mu
    - \gamma\big( B^\mu \,\nabla_\rho B^\rho + \nabla^\mu X \big) \notag \\[6pt]
  &\quad + \lambda\,R\,B^\mu
    - 2\lambda\,R^\mu{}_{\!\nu}\,B^\nu .
\end{align}
\textbf{Metric Field equations}
\begin{equation}\label{21}
    M_{*}^2 G_{\mu\nu} = T_{\mu\nu}^{\rm(eff)}.
\end{equation}
Here,
\begin{align}\label{22}
T^{\rm (eff)}_{\mu\nu} 
   &= \tfrac{1}{2}\,T^{\rm em}_{\mu\nu} \notag\\[6pt]
   &\quad - \tfrac{m^2}{2}\Big[ B_\mu B_\nu 
      - \tfrac{1}{2} g_{\mu\nu} B_\alpha B^\alpha \Big] \notag\\[6pt]
   &\quad - \tfrac{\lambda}{2}\Big[ g_{\mu\nu} R_{\rho\sigma} B^\rho B^\sigma 
      + B_\mu B_\nu R \Big] \notag\\[6pt]
   &\quad + \tfrac{\gamma}{2}\Big[ B_\mu B_\nu \nabla_\rho B^\rho 
      - g_{\mu\nu} B^\rho \nabla_\rho X \Big] \notag\\[6pt]
   &\quad - \tfrac{1}{2} g_{\mu\nu}\, \nabla_\alpha \nabla_\beta A^{\alpha\beta}. 
 \end{align}
and the effective mass 
\begin{equation}
    M_{*}^2=\frac{M_p ^2}{2} -\lambda X;\quad A^{\alpha\beta} = [M_{*}^2g^{\alpha\beta}-\lambda B^\alpha B^\beta].
\end{equation} 

The last term in Eq.\eqref{22} is purely a result of bulk contributions from the term $A^{\alpha\beta} \delta R_{\alpha\beta}$. We can avoid them as done in ref. \cite{Weinberg:1972kfs} by assuming the second-order
derivatives of the metric are zero at the boundary. However, these terms are of no trouble. In the limit $\gamma,\lambda,m,\alpha\rightarrow0$, the theory reduces to Einstein's GR. In this section, we haven't discussed the signature of the coupling constants in detail because it has direct consequences for the stability of the theory, and therefore, it requires special attention, as given in Section \ref{section4}.

\section{Dark energy dynamics}\label{section3}

The cosmological ansatz forces the vector field to be timelike $B^\mu = (\psi(t),0,0,0)$ in the preferred FRLW background,
\begin{equation}
    ds^2 = -dt^2+a^2(t)\delta_{ij}dx^idx^j.
\end{equation}
The vector field equation as shown in Eq.\eqref{20} is non-dynamical in the FRLW background with this cosmological ansatz, but sets up an algebraic relation  of the time-like component $\psi$ with the Hubble parameter $H$ as
\begin{equation}\label{25}
    \psi = \frac{[6\lambda H^2 + m^2]}{3\gamma H}.
\end{equation}
As we can see, $\psi$ is proportional to $H$ and $H^{-1}$. For the late-time cosmological epoch, dominated by $1/H$, sets an observational constraint on the ratio $m/\gamma$ as  $10^{-15}M_p\leq m/3\gamma<15M_p$ \cite{De_Felice_2017} which would prove to be crucial in bounding the parameter space spanned by $\lambda,\gamma,\&\ m$. Furthermore, the modified Friedmann equations for the effective energy density \cite{heisenberg2017generalisedprocatheories} are then the algebraic relations in $H$ and $\psi$;
\begin{align}\label{26}
    T^{\rm (eff)}_{00}=
    &\rho = \frac{1}{2}m^2\psi^2   -3\gamma H\psi^3+\\ \notag&12\lambda H^2\psi^2 + 3M^2_pH^2\\ \notag &-3\lambda\psi^2H^2.
\end{align}
using the right-hand side of Eq.\eqref{21} as 
\begin{equation}
    \rho = 3M_{*}^2 H^2.
\end{equation}
with Eq.\eqref{26} and Eq.\eqref{25} giving a cubic equation in $H^2$ providing roots $(H_*^2)$ which are the stationary points as shown below
\begin{equation}\label{28}
324 \lambda^{3} H^{6} 
+ \big(72 \lambda^{2} m^{2} + 27 \gamma^{2} M_{p}^{2}\big) H^{4} 
- 3 \lambda m^{4} H^{2} 
- m^{6} = 0.
\end{equation}

The physical viability of this equation requires at least one root to be positive, that is, $H_*^2>0$. According to Descartes' rule of signs \cite{bensimhoun2016historicalaccountultrasimpleproofs}, the sign pattern of our polynomial Eq.\eqref{28} $(+,+,-,-)$ shows one sign flip, which confirms at least one positive root for $\lambda>0$. However, for $\lambda<0 $ the sign pattern of the polynomial becomes $(-,+,+-)$, hinting at the presence of either two or zero positive roots. Further analysis confirms the presence of two positive roots, as the slope with respect to $H^2$ is positive at $H^2 = 0$. This seems to have a richer phenomenology, but it requires stability analysis to confirm the signature of $\lambda$. The discriminants of both configurations are positive; hence, all the roots are real and distinct, given by the formula of François Viète \cite{Nickalls_2006}.   
\begin{align}\label{29}
    H_*^2(k)=2\sqrt{\frac{-p}{3}}\cos\big[\frac{1}{3}\arccos\Big(\frac{3q}{2p} \sqrt{\frac{-3}{p}}\Big)
    -\frac{2k\pi}{3}\big]\\\notag
    \quad-\frac{8 \lambda^{2} m^{2} + 3 \gamma^{2} M_{p}^{2}}{108 \lambda^{3}}.
\end{align}
for $k=0,1,2$.
(see the appendix for the explicit expressions for $p$ and $q$) We cannot comment analytically about the value of $k$ for which one of the three roots will be positive; one needs numerical analysis. We can establish an inequality here for Eq.\eqref{29} to be positive, that is for the positive maximum $\cos(....)$ it is naturally imposed that,
\begin{equation*}
    2\sqrt{\frac{{-p}}{3}} > \frac{8 \lambda^{2} m^{2} + 3 \gamma^{2} M_{p}^{2}}{108 \lambda^{3}}.
\end{equation*}
Without commenting on the sign of $\lambda$ for a short time, we must proceed to another important equation in the path of calculating the EoS parameter for the theory.  The effective pressure for the metric field equations is  given by;
\begin{align}
g^{ij} T_{ij}^{\rm (eff)} &= -P \\\notag
&= -\frac{1}{2} m^2 \psi^2 
   - \gamma \dot\psi \psi^2 
   + [M_p^2 - \lambda \psi^2] [3 H^2 + 2 \dot H] \\ \notag
&\quad + 2 \lambda \psi [3 H^2 \psi + 2 H \dot\psi + 2 \dot H \psi].
\end{align}
along with the right hand side of the equation Eq.\eqref{21} $-P = M_*^2[3H^2+2\dot H]$ with, $u = \dot H/H^2$. We have an ordinary differential equation for $H(t)$, 
\begin{equation}\label{31}
    g(H) u +f(H) = 0 ,
\end{equation}
\begin{equation}
\begin{aligned}\label{32}
f &= -9 H^{2} \lambda m^{4} - 3 m^{6}  \\
  &\quad + 216 \lambda^{2} m^{2} H^{4} 
         + 81 \gamma^{2} M_{p}^{2} H^{4} 
         + 972 \lambda^{3} H^{6},
\end{aligned}
\end{equation}
\begin{equation}
\begin{aligned}
g &= 6 H^{2} \lambda m^{4} + 2 m^{6} \\
  &\quad + 144 \lambda^{2} m^{2} H^{4} 
         + 54 \gamma^{2} M_{p}^{2} H^{4} 
         + 1080 \lambda^{3} H^{6}.
\end{aligned}
\end{equation}
We do not need to solve the equation Eq.\eqref{31} for the explicit form of $H(t)$ to evaluate the EoS parameter $w$, as it is simply $w=-1-2u/3$. Substituting $u$ from Eq.\eqref{31} in terms of $H$, $w$ takes the algebraic form in $H$, 
\begin{equation}\label{34}
w=-\frac{2 \left(108 H^{6} \lambda^{3} + 3 H^{2} \lambda m^{4} + m^{6}\right)}
{540 H^{6} \lambda^{3} + 3 H^{2} \lambda m^{4} + m^{6} + 9 H^{4} \left(8 \lambda^{2} m^{2} + 3 \gamma^{2} M_{p}^{2}\right)}.
\end{equation}
The moment we decouple the Ricci scalar and tensor from the vector field, that is, setting $\lambda =0$, the EoS parameter $w$ becomes,
\begin{equation}\label{35}
    w= -\frac{2m^6}{m^6+27\gamma^2M_p^2H^4}.
\end{equation}
which is close to $\Lambda$CDM, i.e., $w\sim-1$, for the order $m^6\sim27\gamma^2M_p^2H^4$ at the current value of Hubble parameter $H_0$. Having the EoS in hand, Eq.\eqref{34}
We can analyze it for both the branches of $\lambda$. Let the numerator and denominator be;
\begin{align}\notag
    &N(y) = 108y^3 \lambda^{3} + 3 y \lambda m^{4} + m^{6},\\\notag
    &   D(y) = 540 y^3 \lambda^{3} + 3 y \lambda m^{4} + m^{6} + 9 y^2 \left(8 \lambda^{2} m^{2} + 3 \gamma^{2} M_{p}^{2}\right) .
\end{align}
For $\lambda>0$ in the region $y=H^2>0$ numerator is always positive as the sign pattern is $(+,0,+,+)$, and the absence of any sign flip prevents the factorization of $N(y)$ in the physically viable region $y>0$. At the same time, the denominator $D(y)$ is always positive as the sign pattern is $(+,+,+,+)$ and there is no sign flip, hence no factorization in the viable region. This clearly shows that for $\lambda>0$ there are no pathologies in $w$. Both the numerator and denominator in the Negative branch $\lambda<0$ are factorizable. For the numerator, the sign pattern $(-,0,-,+)$ with exactly one sign flip hints at exactly one positive solution for $N(y) = 0$. For the denominator, the sign pattern is $(-,+,-,+)$ strongly hinting at the presence of 3 or 1 positive real solutions for $D(y) =0$ (\textit{The cubic discriminant of $D(y)$ takes the negative value, hence only one real positive root).} One can easily see that there is no common root that can be factored out and prevent the $w$ from blowing to $-\infty$ during the viable cosmological epoch $y>0$. To avoid this kind of pathology, it is safe to consider the positive branch as a physically viable branch. The signature of $\gamma$ neither affects the EoS nor the value of stationary solutions Eq.\eqref{29} like $\lambda$. Additionally, it does not affect the null energy condition Eq.\eqref{39} and the condition on the perturbative stability against scalar ghost Eq.\eqref{40}. It only affects the signature of the time-like component $\psi$ of the vector field and its derivative, as one can see from equation Eq.\eqref{25}. Mathematically, both $(\gamma,-\gamma)$ are equivalent, in cosmology $\psi>0$ such that the timelike vector aligns with the comoving observer's direction of time. Hence, we prefer $\gamma$ to be positive. We are then forced to consider $m^2 > 0$ to avoid tachyonic instabilities by construction. To check whether our stationary solution Eq.\eqref{29} is a de-sitter attractor or a repellor using Lyapunov's second method for stability\cite{lyapunov1992general}, treating $V(H) = (H-H_*)^2/2$ as the Lyapunov function, with $\dot H = -H^2f(H)/g(H)$ from Eq.\eqref{31}, we require $\dot V<0$. Note that at the stable point $f(H_*) =0 , \dot H=0$, using Eq. (\ref{31}).
\begin{equation}\notag
    \dot V = \frac{\partial V}{\partial H} \dot H = -(H-H_*)\frac{H^2f(H)}{g(H)}.
\end{equation}
Linearizing $f(H)$ around $H_*$ as $f(H)\sim(H -H_*)f'(H)\Big|_{H = H_*}$ finally we have,
\begin{equation}\notag
    \dot V = \frac{\partial V}{\partial H} \dot H = -(H-H_*)^2\frac{H^2f'(H_*)}{g(H)}.
\end{equation}
Since $g(H)$ is always positive for $(H-H_*)^2>0$, and $H^2$ is positive too, the sign of the overall expression depends on the sign of $f'(H_*)$. It is easy to see from the expression for $f(H^2)$ Eq.\eqref{32} that $f(0) = -3m^6$ and $f(\infty) = \infty$ with only one positive root in $H^2>0$ one can conclude that the polynomial crosses the divide $f(H^2) = 0 $ from negative to positive at $H= H_*$ that is $f(H^2)<0$ to $ f(H^2)>0$ confirming $f'(H_*) >0$ making over all, $\dot V<0$. Hence, the stationary point is de-sitter attractor. Analytically, we have proved $f'(H_*)>0$ algebraically $f'(H_*) \propto H_* \cdot (\text{Polynomial in } H_*^2$) . For $f'(H_*)>0$ clearly there are two cases 1. $H_*, (\text{Polynomial in } H_*^2 )\quad \text{both} >0$ and 2. both $<0$  the second case refers to the contracting de Sitter spaces that are unstable with respect to tensor perturbations as shown in \cite{Faraoni_2004}, hence we are forced to consider $H_*>0$, which further helps to conclude that the point $H_*$ is a de-sitter attractor.

\section{Stability and Constraints}\label{section4}
We ensured that our theory Eq.\eqref{2} is free from Ostrogradsky instabilities by embedding it within the GP framework. However, this alone is not sufficient---a full perturbative analysis is required to verify the absence of scalar, vector, and tensor ghosts, as well as gradient instabilities, as thoroughly examined in \cite{De_Felice_2016}. Furthermore, beyond the stability requirements, the parameter space spanned by $\lambda$, $\gamma$, and $m$ must also be constrained using observational data, as discussed in this section. All the stability conditions and the observational constraints mentioned below isolate the physically viable region in the parameter space.

\bigskip

\noindent\textbf{Stability constraints}

\bigskip

\noindent\textbf{1.} Tensor ghosts and small-scale Laplacian instabilities are absent for $c_\tau^2>0$ which implies for our case;
\begin{equation}\notag
    2G_4>0, \quad\& \quad2G_4+2\lambda\psi^2>0.
\end{equation}
for which we can say;
\begin{equation}\label{36}
    -\frac{M_P^2}{\psi^2}<\lambda<\frac{M_P^2}{\psi^2}.
\end{equation}\\
\noindent\textbf{2.}The small-scale stability from the vector mode propagation speed $c_v^2>0$,
\begin{equation}\label{37}
    c_v^2 = 1+\frac{\psi^2\lambda^2}{M_p^2+\lambda\psi^2} > 0.
\end{equation}
This condition is inherently satisfied when analyzed with Eq. (\ref{36})

\bigskip

\includegraphics[width=9cm]{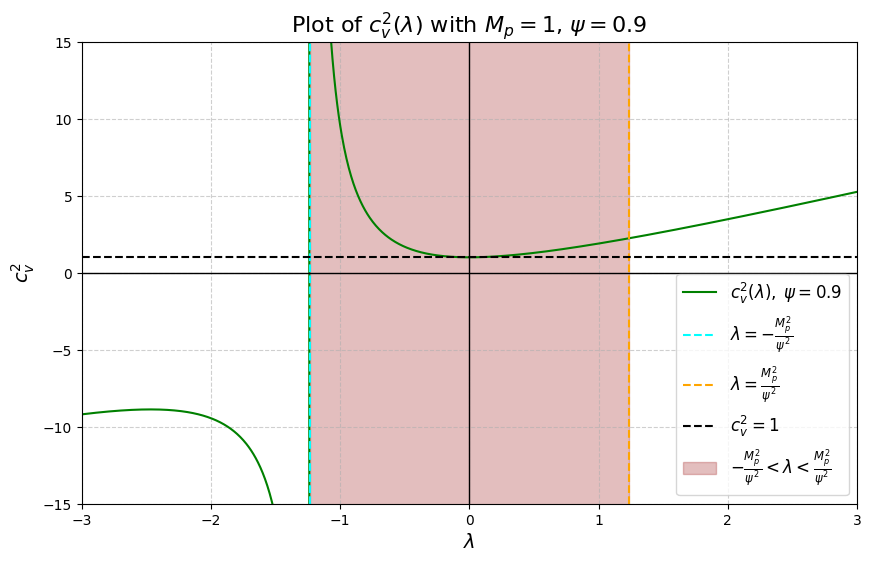}
The figure above shows that the vector speed is superluminal and blows to infinity for $\lambda \to -M_p^2/\psi^2$, whereas for the positive limit, $c_v^2$ is finite. We can safely ignore the negative branch from the arguments made in the previous section about the sign of $\lambda$. This sets an upper limit on the vector mode propagation as $\lambda\to M_p^2/\psi^2$, which is;
\begin{equation}
    \lim_{\lambda\to M_p^2/\psi^2}c_v^2  = 1+\frac{M_p^2}{2\psi^2}.
\end{equation}
For small values of $\lambda$, the speed of propagation for vector modes is almost the speed of light $c_v^2\sim1$

\smallskip

\noindent\textbf{3.} For the current value of the Hubble constant $H_0$, the null energy condition \cite{hawking1973large} $ P+\rho>0$ must be held true.
\begin{center}
\begin{align}\label{39}
     \frac{M_{*}^2H^2f}{g}>0 \quad\text{or} \quad f(H)>0.
\end{align}
\end{center}
Whereas for the stationary point $H_*$, the condition $p+\rho =0$ is satisfied canonically. As discussed in Section \ref{section3}, for only one stable stationary solution $H_{*}$ , $f'(H_{*})>0$. Hence for $H>H_{*}$, $f(H)$ is always positive. Which ultimately respects the null energy condition.

\bigskip

\noindent\textbf{4.} The absence of scalar ghosts demands, 
\begin{equation}\label{40}
    48\lambda^2\psi H^2+(6\gamma M_p^2-18\gamma\lambda\psi^2)H+3\gamma^2\psi^3 >0.
\end{equation}
This equation is verified as a true statement once we substitute \ref{25}.
\begin{equation*}
    \frac{48 H^{3} \lambda^{3}}{\gamma} + 6 H M_{p}^{2} \gamma + \frac{4 H \lambda^{2} m^{2}}{\gamma} + \frac{m^{6}}{9 H^{3} \gamma}>0.
\end{equation*}
Since all the parameters are positive, the inequality holds, indicating the absence of scalar ghosts.

\bigskip
 
The above conditions collectively ensure the theoretical stability of the model. However, these analytical constraints alone are insufficient to tightly restrict the parameter space. To obtain a physically viable region, one must further incorporate numerical bounds derived from observational data on cosmological observables.
 
\bigskip

\noindent\textbf{Observational Constraints:}

\bigskip

\noindent\textbf{1.} The cosmological variation of the effective gravitational constant \( G_{\rm eff} \) \cite{Uzan_2003} is subject to a stringent upper limit,
\begin{equation}\label{41}
    \Bigg|\frac{\dot{G}_{\rm eff}}{G_{\rm eff}}\Bigg| \leq 10^{-12}\, \text{year}^{-1}.
\end{equation}
The explicit expression for $G_{\rm eff}$ in the present framework is provided in Appendix  \ref{appendix}. Although this constraint has not been employed here to delimit the parameter space due to the highly nonlinear dependence of $G_{\rm eff}$ on the model parameters, it is important to note that such a bound can, in principle, impose a much tighter restriction on the viable parameter region. Hence, it is included for completeness and to highlight its potential role in a more comprehensive analysis.

\smallskip

\noindent\textbf{2.} The propagation speed of gravitational waves predicted by this model is subluminal and is tightly constrained by the stringent Cherenkov radiation bound \cite{Moore_2001, LIGOScientific:2017zic},
\begin{equation}
    1 - c_\tau \leq 2 \times 10^{-19}.
\end{equation}
In our framework, the tensor propagation speed is given by
\begin{equation}
    c_\tau^2 = \frac{F_T}{G_T} = \frac{M_p^2 - \lambda\psi^2}{M_p^2 + \lambda\psi^2}.
\end{equation}
Applying the stability requirement \( M_p^2 > \lambda\psi^2 \) for \( \lambda > 0 \), and evaluating at the present Hubble rate \( H = H_0 \), we obtain
\begin{equation}\label{44}
    \Bigg|\frac{\lambda \psi^2}{M_p^2}\Bigg| \leq 2 \times 10^{-19}.
\end{equation}
This observational bound provides a crucial constraint on the model parameters, effectively placing a strict upper limit on $\lambda $. Owing to its direct connection with gravitational-wave observations, this condition serves as one of the most significant empirical checks on the theoretical consistency of the model.

\bigskip

\noindent\textbf{3.} In the quasi-static (post-Newtonian) limit of the theory, relevant at Solar System scales, one can safely assume $\dot{\psi} \simeq 0 $. Under this approximation, the post-Newtonian parameter $\gamma_{\text{PPN}} $ (as defined in ref.\cite{Will1981}) provides an additional constraint on the model parameters:
\begin{equation}\label{45}
    \gamma_{\text{PPN}} = \frac{M_p^2 - \frac{\lambda}{2}\psi^2}{M_p^2 + \frac{\lambda}{2}\psi^2 + 4\lambda^2\psi^2}.
\end{equation}
The Cassini Shapiro time-delay experiment~\cite{Bertotti:2003rm} imposes a stringent observational bound,
\begin{equation*}
    |\gamma_{\text{PPN}} - 1| \in [0.2 \times 10^{-5},\, 4.4 \times 10^{-5}].
\end{equation*}
Although this constraint is included here for completeness, it is not employed in the present analysis to delimit the parameter space. Enforcing the Cassini bound in conjunction with the Cherenkov constraint on the tensor propagation speed [Eq.\eqref{44}] leads to an extremely high upper limit on the coupling parameter, $ \lambda \sim \mathcal{O}(10^{13}) $. Such a value renders the inverse mapping $ H(\psi)$ of $ \psi(H)$, derived from Eq.\eqref{25}, imaginary when evaluated together with the observationally bounded values of the ratio $m / 3\gamma $. This behavior indicates that these large values of $\lambda$ lie outside the physically admissible regime of the model, implying that the Solar System constraint alone cannot be relied upon to establish meaningful upper bounds on the parameters.

\bigskip

The observational constraint in Eq.\eqref{44} thus represents a more stringent and physically viable version of the tensor ghost stability condition in Eq.\eqref{36}. Moreover, since the vector mode stability condition [Eq.\eqref{37}] is automatically satisfied whenever Eq.\eqref{36} holds—and the latter is guaranteed by the observational bound in Eq.\eqref{44}.

\section{Results}\label{section5}

Having established the theoretical consistency conditions and observational constraints governing the parameter space, we now proceed to examine the viable regime of the model by rescaling all dimensional quantities with respect to the current Hubble parameter $H_0$. This normalization not only simplifies the numerical analysis but also facilitates a direct comparison with present cosmological observations. The rescaled parameters are defined as
\begin{align*}
     &m = \tilde mH_0, \\
     &\gamma = \tilde \gamma H_0/M_p, \\
     &H = \tilde HH_0, \\
     &\psi(H) = \tilde\psi(\tilde H) M_p.
\end{align*}
The observational constraint on the mass-to-coupling ratio discussed earlier is accordingly expressed as 
\[
10^{-13}\leq \frac{\tilde m}{3\tilde \gamma}<15,
\]
where we define the dimensionless ratio $\tilde m / 3\tilde\gamma = r$. Consequently, the $(m,\gamma)$ parameter plane is effectively bounded by the domain of $r$. Meanwhile, the upper limit on the remaining coupling parameter $\lambda$ arises from two independent requirements: (i) the Cherenkov bound on the propagation speed of gravitational waves, and (ii) the condition for the existence of real roots when inverting Eq. (\ref{25}), that is, $\tilde\psi(\tilde H) \to \tilde H(\tilde\psi)$. The determinant associated with this inversion,
\[
9\tilde\gamma^2\psi^2 - 24\lambda \tilde m^2,
\]
must remain positive at the present epoch ($\tilde H = 1$), imposing a lower bound on $\tilde\psi^2$ as
\[
\tilde\psi^2(\tilde H=1) > \frac{8\lambda \tilde m^2}{3\tilde\gamma^2}.
\]
Combining this condition with the upper bound from the energy density constraint yields
\[
\frac{8\lambda \tilde m^2}{3\tilde \gamma^2}<\tilde \psi^2(\tilde H=1) \leq \frac{2\times 10^{-19}}{\lambda}.
\]
Following from this logical relation, one obtains
\[
\frac{8\lambda \tilde m ^2}{3\tilde \gamma^2}<\frac{2\times 10^{-19}}{\lambda} \quad \text{or} \quad \frac{8\lambda^2 \tilde m ^2}{3\tilde \gamma^2}<2\times 10^{-19}.
\]
Given that the ratio $r$ is bounded numerically, the upper limit on $\lambda$ is then derived as
\[
\lambda<\sqrt{\frac{2\times10^{-19}}{r_{\text{max}}^2}},
\]
which evaluates to $\lambda<2.98 \times 10^{-11}$ for $r_{\text{max}} =15$.

Having determined the permissible bounds for the rescaled parameters, we now measure the deviation $\delta$ at the current epoch, defined in terms of the rescaled couplings. The corresponding EoS takes the form
\begin{equation*}
    w(\tilde H) = -1+\delta, \quad \text{for} \quad \delta = \tfrac{2f(\tilde H)}{3g(\tilde H)}.
\end{equation*}
\begin{figure}[H]
    \centering
    \includegraphics[width=9cm]{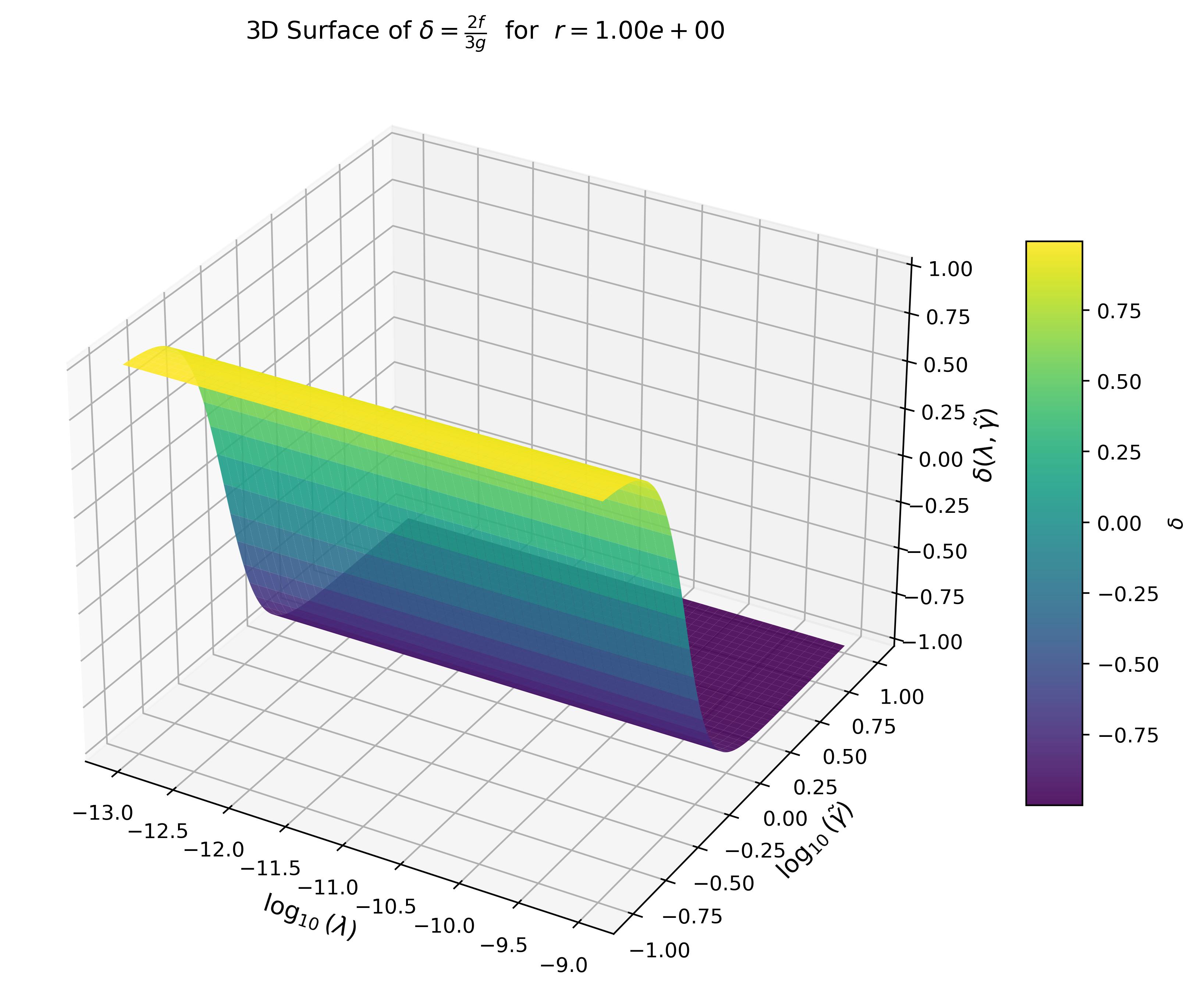}
    \caption{Deviation ($\delta$) of the equation-of-state parameter $w(\tilde H)$ for the allowed range of model parameters.}
    \label{Figure 1}
\end{figure}
The figure above illustrates the variation of the deviation parameter $\delta$ in the EoS, corresponding to the free parameters $\lambda$ and $\tilde\gamma$, evaluated at the present Hubble scale ($H = H_0$ or equivalently $\tilde H = 1$). The colored projection in the $\log$–$\log$ plane is displayed below:
\includegraphics[width=9cm]{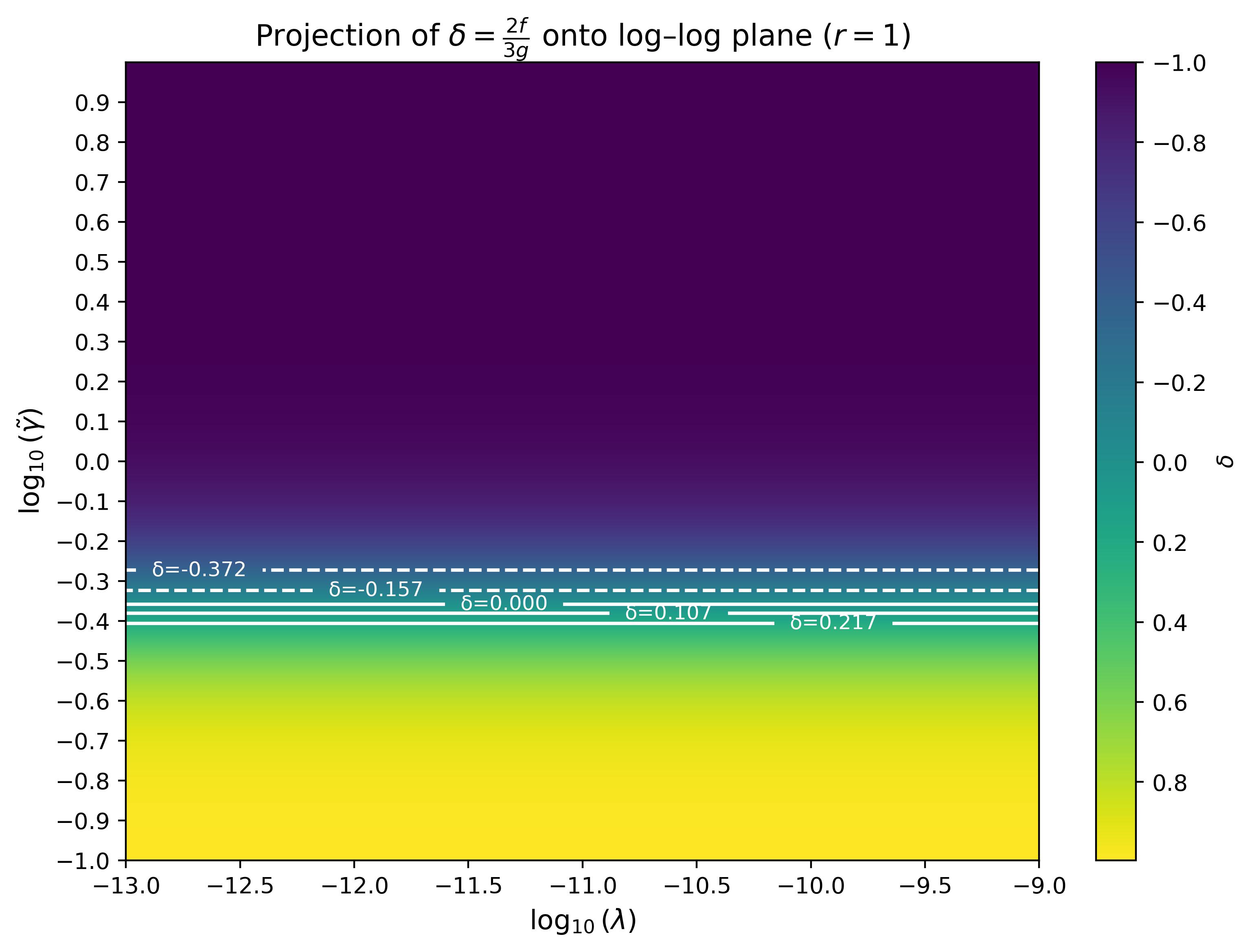}\label{figure 3}
Upon close inspection, one finds that within the band $-0.405 \geq \log_{10}\tilde\gamma \geq -0.38$, the deviation $\delta$ lies in the range $0.107 \leq \delta \leq 0.217$. This range successfully captures the deviation of the present-day EoS parameter $w_0 = w(\tilde H = 1)$ inferred from recent DESI + CMB + Pantheon+ observations \cite{DESI:2025zgx}.
\begin{figure}[H]
    \centering
    \includegraphics[width=9cm]{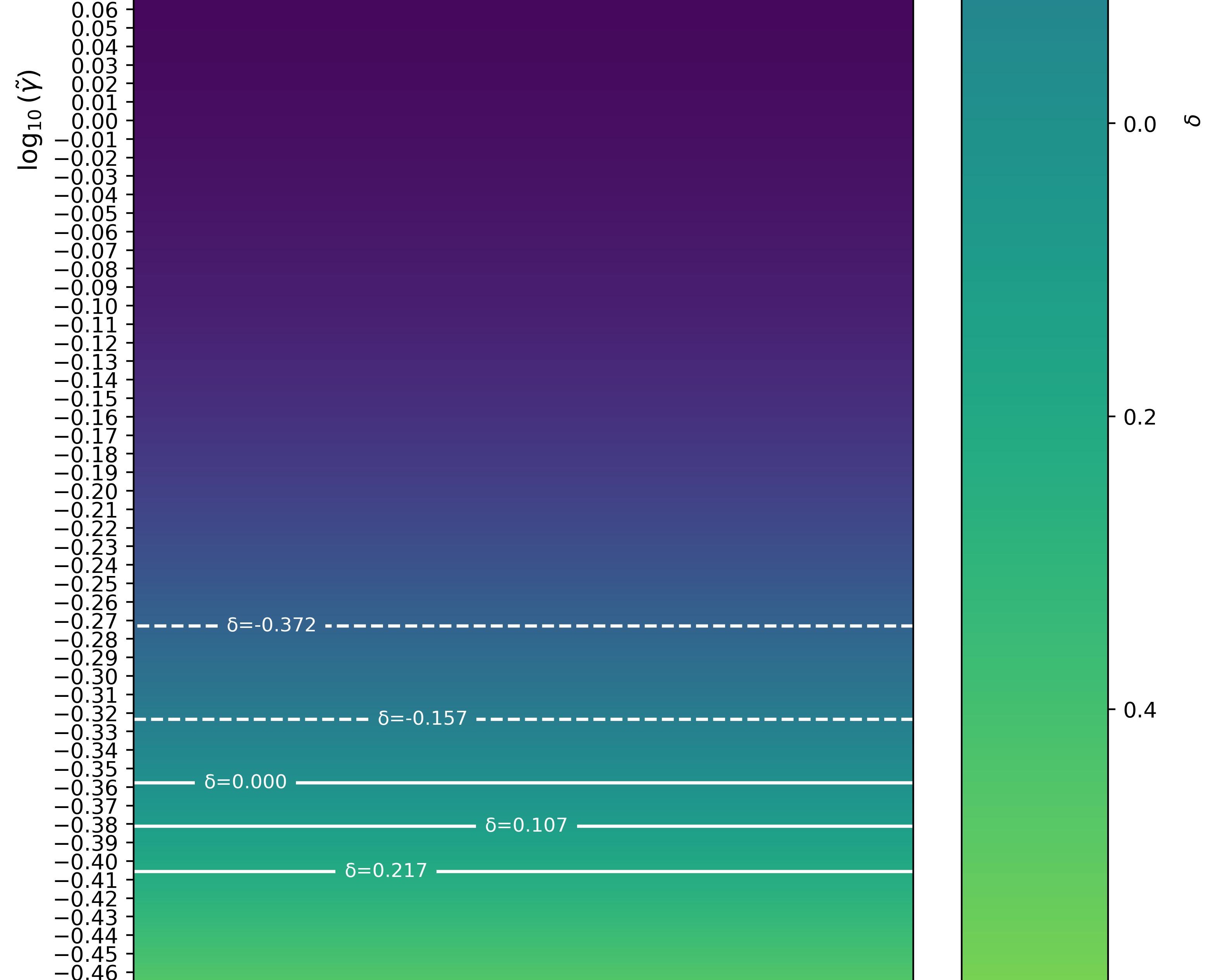}
    \caption{Zoomed-in projection highlighting the viable range of $\tilde\gamma$ consistent with the observational deviation $\delta$ from DESI+CMB+Pantheon+ data.}
    \label{figure 2}
\end{figure}
The viable range of $\tilde\gamma$ obtained from this analysis is also consistent with the observational upper limit on the photon mass $m$ derived from independent measurements such as localized Fast Radio Bursts (FRBs)\cite{Ran:2024avn,Wang:2023fnn} and rotating torsion balance experiments \cite{Luo:2003rz}, which yield $m \leq 3\times10^{-48}\text{g}$. Within the identified interval $-0.405 \geq \log_{10}\tilde\gamma \geq -0.38$, the corresponding mass range is 
\[
1.18065H_0 \geq m \geq 0.41686H_0,
\]
which translates to
\[
3.1356\times10^{-66}\text{g} \leq m \leq 3.3627\times10^{-66}\text{g},
\]
for $H_0 = 70\text{km s}^{-1}\text{Mpc}^{-1}$. The lower bound on $m$ can be interpreted as a hint from the DESI DR2 observations indicating possible evidence for a dynamical dark energy component.

For ratios $r \leq 10^{-4}$, the deviation reaches $\delta = 1$, which corresponds to $w_0 = 0$ and thus violates the current observations within the given parameter space, as illustrated in the following figure:
\begin{figure}[H]
    \centering
    \includegraphics[width=9cm]{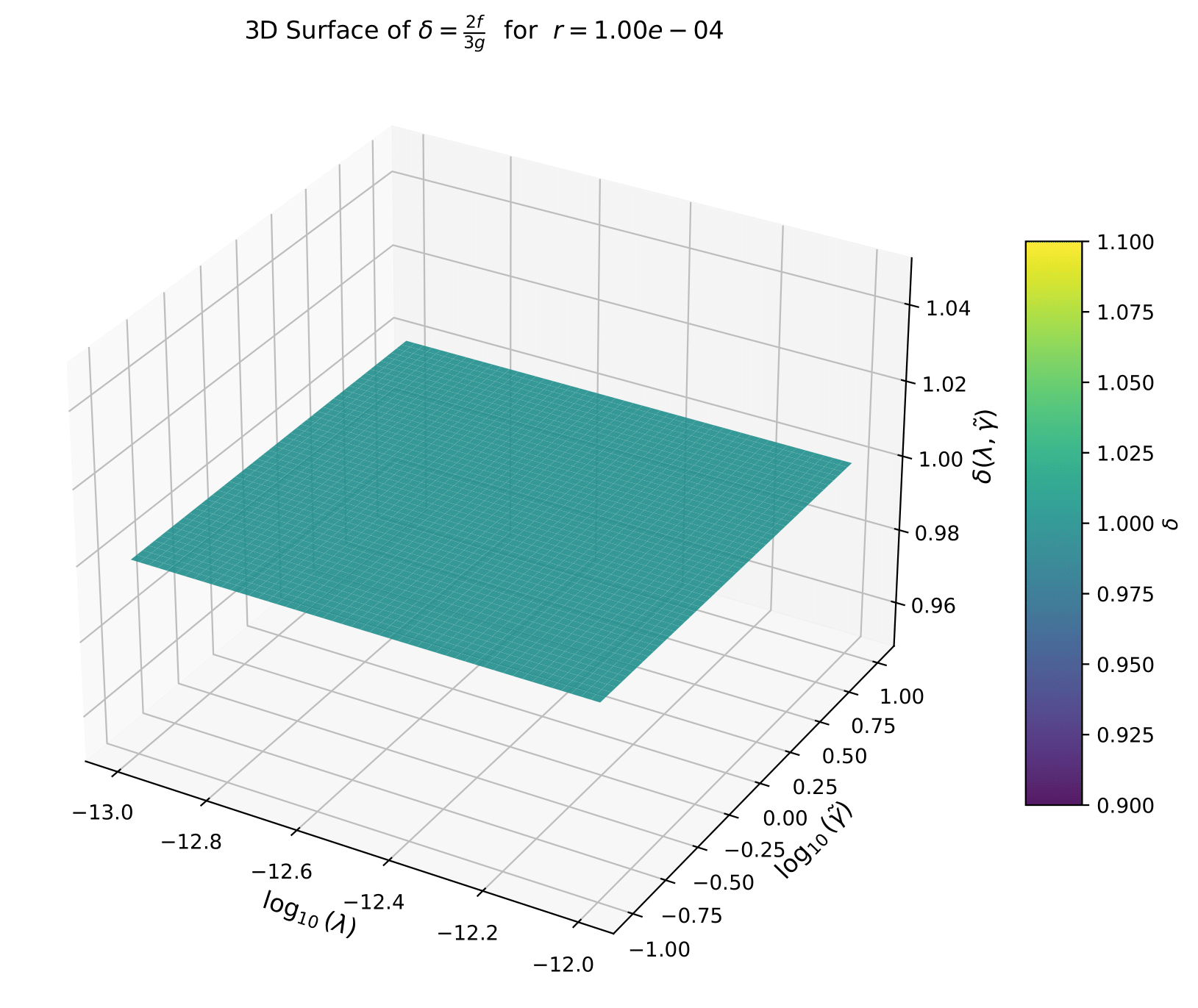}
    \caption{Variation of $\delta$ with respect to $r$ showing the unphysical region ($\delta=1$ corresponds to $w_0=0$).}
    \label{Figure 2}
\end{figure}
From Figure \ref{Figure 1} and its corresponding projections, we observe that an increase in the coupling strength $\tilde\gamma$, which quantifies the self-interaction of the field, drives the system into the phantom regime ($w_0 < -1$). Conversely, a decrease in $\tilde\gamma$ leads to a matter-like phase ($w_0 \sim 0$). Therefore, the strength of the self-interaction must remain within the narrow viable range identified above to maintain consistency with cosmological observations.
For these physically admissible couplings, the behavior of $w(\tilde H)$ as a function of the rescaled Hubble parameter is shown below:
\begin{figure}[h!]
    \centering
    \includegraphics[width=\columnwidth]{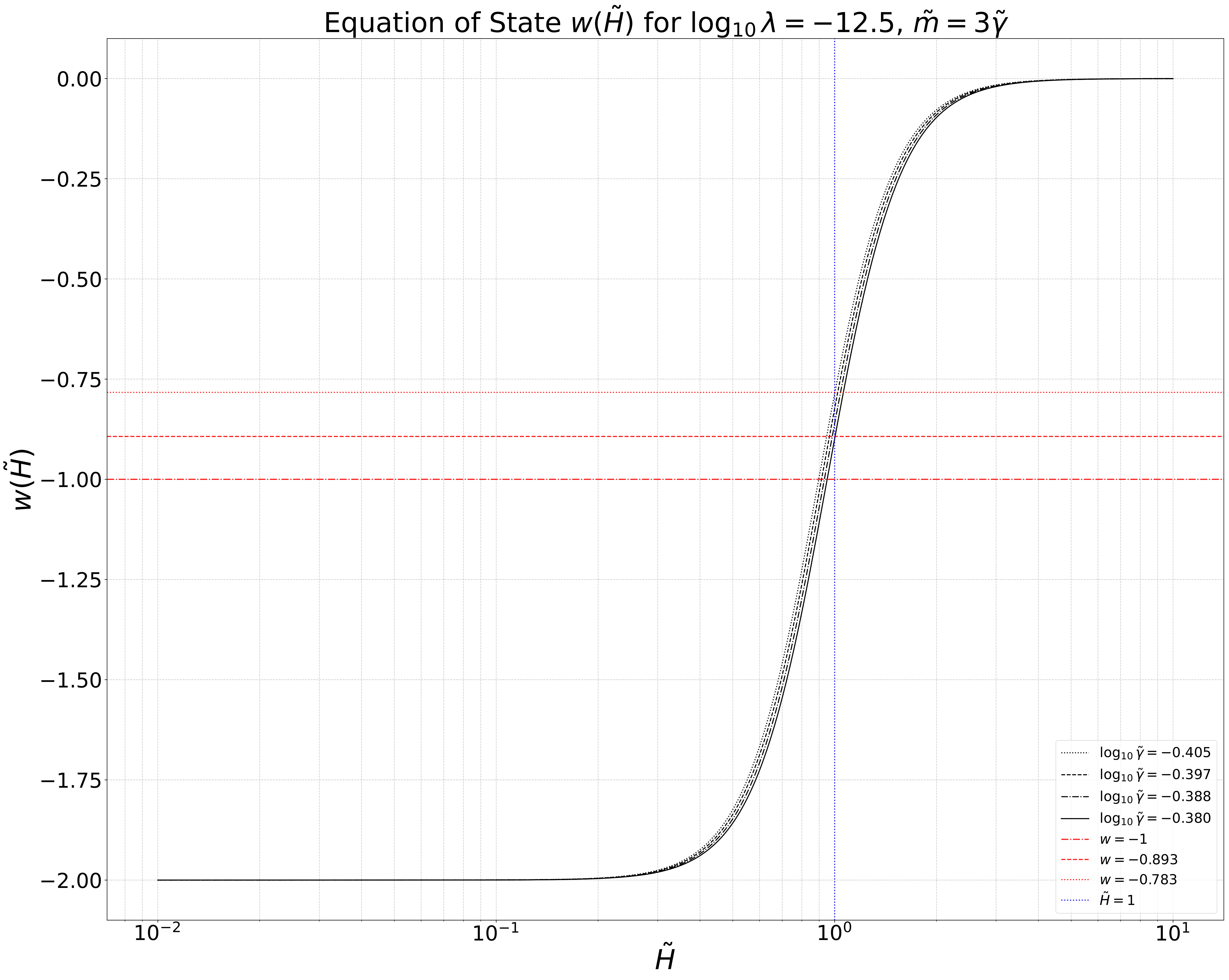}
    \caption{EoS Vs Rescaled Hubble parameter for different values of $\tilde \gamma$.}
    \label{a}
\end{figure}
From the above plot, it is evident that the stationary point $\tilde H_* $, for which $w(\tilde H_*) = -1$, lies to the left of the current Hubble scale ($\tilde H = 1$). This feature confirms the existence of a single stationary de Sitter attractor, situated near the present epoch. 
\includegraphics[width=9cm]{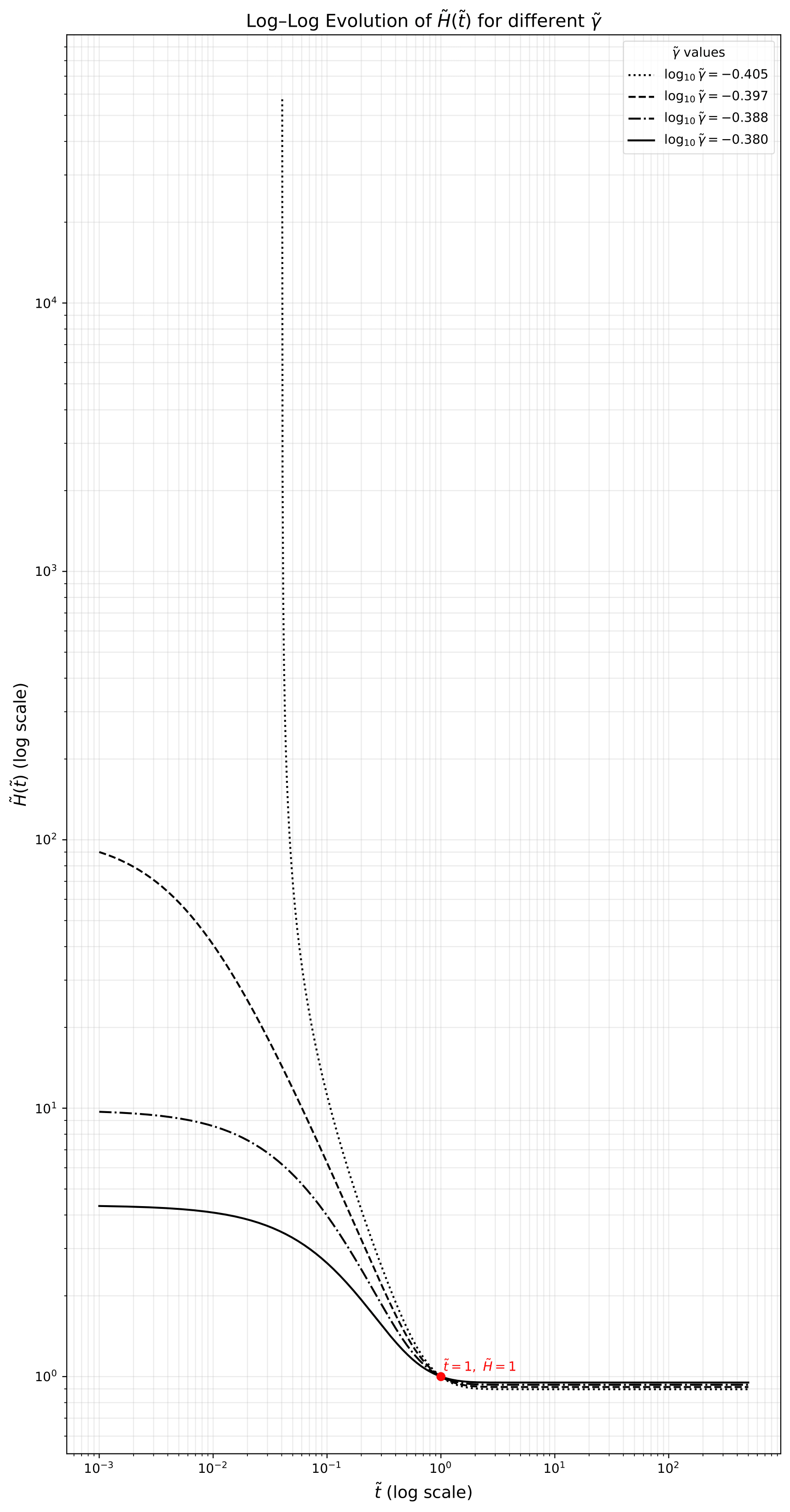}
The above plot illustrates the evolution of the rescaled Hubble parameter, $\tilde{H} = H/H_0$, as a function of the rescaled cosmic time, $\tilde{t} = tH_0$ governed by the differential equation \eqref{31}, where $\tilde{H} = 1$ and $\tilde{t} = 1$ correspond to the present-day Hubble constant and the current age of the Universe, respectively. Upon zooming into the vicinity of the present epoch, as shown in the figure below, one observes that the Hubble parameter asymptotically approaches a stationary de Sitter value. For instance, at $\tilde{t} \gtrsim 2$ (equivalently, $t \gtrsim 28,\mathrm{Gyr}$ as shown in Fig. \ref{a}), the late-time expansion settles into a phase characterized by the EoS $w = -1$. Notably, the model remains on the non-phantom side, never crossing the $w = -1$ divide.
\includegraphics[width=9cm]{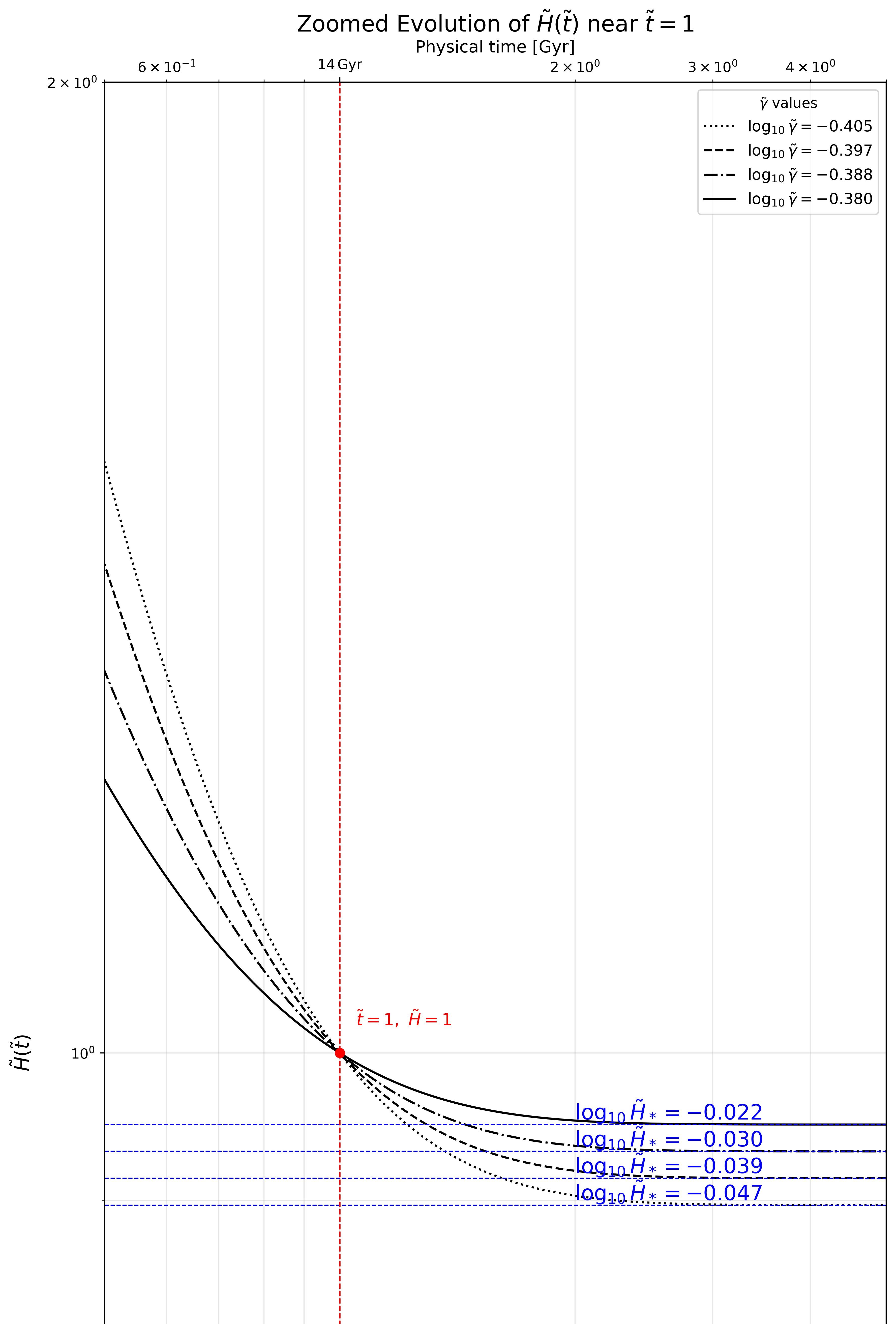}

\section{Discussions and Conclusion}\label{section6}
In Section II, we embedded our Lagrangian into the Generalized Proca (GP) framework to ensure its stability against Ostrogradsky ghost degrees of freedom, an aspect that was not addressed in the earlier work \cite{B_hmer_2007}. The resulting Lagrangian, prior to integrating out the boundary contributions given by Eq.\eqref{19}, indicates that for the theory described by Lagrangian in Eq. (\ref{2}) to remain stable, the couplings between the Ricci scalar and the divergence of the vector field, as well as those between the Riemann tensor and the derivative of the vector field, must be arranged such that their net contribution manifests purely at the boundary, leaving the bulk free of ghost instabilities.

\bigskip

However, these boundary terms are far from trivial. In light of the Holographic Principle \cite{10.1063/1.531249,RevModPhys.74.825}, which states that \textit{a higher-dimensional physical theory (the bulk) can be completely described by a theory defined on its lower-dimensional boundary}, such contributions play a critical role. Specifically, the terms $\alpha G_{\mu\nu}\nabla^\mu B^\nu$ and $M_*^2 R$ contribute to the Wald entropy~\cite{Conroy_2015,Wald_1993,Brustein_2011}, leading to a natural deviation from the standard Bekenstein–Hawking area law $S \propto A$~\cite{PhysRevD.7.2333}. Consequently, one can predict the fate of the universe by this approach and compute the associated Noether charges within the GP framework.

\bigskip

In Section \ref{section3}, the cosmological ansatz constrains the field evolution to follow Eq.\eqref{25}, effectively addressing the long-standing issue of working with $\psi^p \propto 1/H$ in marginally coupled theories, as discussed in the introduction. Furthermore, Eq.\eqref{25} assists in determining the upper bound on the relevant coupling $\lambda$, as elaborated in the previous section. The modified Friedmann equation that follows is a sixth-order polynomial in $H$, whose only physical solution corresponds to a stationary de Sitter attractor, as demonstrated therein. We also emphasized that the couplings between the vector field and curvature must remain positive to prevent poles in the EoS during viable cosmological epochs.

\bigskip

In Section \ref{section4}, all necessary stability conditions—including the null energy condition at the present epoch ($\tilde{H} = 1$) were verified. The observational constraint on the propagation speed of gravitational waves subsequently served as the key condition for setting an upper bound on $\lambda$, as shown in Section V. Notably, attempting to restrict the vast landscape of the GP theory by imposing $c_\tau = 1$ (or equivalently $\lambda = 0$) yields an EoS, Eq.(~\ref{35}), that depends explicitly on the vector field mass, the strength of self-interactions ($\gamma$), and the time-dependent Hubble parameter, governed by the differential equation Eq.(\eqref{31}).

\bigskip

Our results in Section \ref{section5} predict an upper bound of $\lambda < 2.98\times10^{-11}$, which is significantly more restrictive than the previously suggested value $\lambda \sim 9000$ reported in Ref.\cite{B_hmer_2007}. This disparity arises primarily from the inclusion of post-Newtonian (PPN) constraints on the cosmological vector field, as discussed in Section \ref{section4}. When the solar-system PPN constraint Eq.\eqref{45} is combined with the strong Cherenkov bound Eq.\eqref{44}, the resulting limit drives $\lambda$ to values on the order of $10^{13}$, rendering the inverse relation $H(\psi)$ imaginary and hence not physical.

\bigskip

Finally, in Section \ref{section5}, we demonstrated that the self-interaction terms associated with the coupling $\gamma$ are the dominant drivers of cosmic expansion, validating the claim made in the introduction. The mass of the cosmological vector field is found to be bounded from below when accounting for the dynamical nature of dark energy, directly due to the constraints on $\tilde{\gamma}$. Strong self-interactions push the universe into the phantom regime ($w_0 < -1$), while weaker interactions correspond to a matter-dominated phase ($w_0 \sim 0$) at the current epoch. This intrinsic duality in the nature of self-interacting terms presents a promising avenue for future investigations—particularly in exploring gravitational collapse models to determine whether strong self-interactions can prevent the formation of a singularity. Notably, the parameter window $-0.405 \leq \log_{10}\tilde{\gamma} \leq -0.38$ corresponds to $0.107 \leq \delta \leq 0.217$, which successfully reproduces the observed present-day EoS $w_0$ within the DESI + CMB + Pantheon+ bounds~\cite{DESI:2025zgx}. The evolution profile of the rescaled Hubble parameter under the viable parameter ranges, as shown in Fig. \eqref{a}, predicts the $\Lambda\text{CDM}$ like behavior for the late cosmic time. This agreement provides a strong observational footing for the proposed framework.
  
\section*{Acknowledgments} 
My sincere thanks go to Mr Ayush Bidlan for his helpful discussions and support in understanding and analyzing the numerical results.

\section*{Appendix}\label{appendix}
\textbf{1.} In regards to Eq. (\ref{29}) the algebraic forms of $p,q$ are
\begin{equation}\notag
    p = -\frac{2916 \lambda ^3 m^4+\left(72 \lambda ^2 m^2+27 \gamma ^2 \text{Mp}^2\right)^2}{314928 \lambda ^4}.
\end{equation}
and,
\begin{align}\notag
q=&\frac{
-2834352 \lambda^{4} m^{6} 
+ 2 \bigl(72 \lambda^{2} m^{2} + 27 \gamma^{2} M_{p}^{2}\bigr)^{3}
}{918330048 \lambda^{6}} \notag \\
&\quad + \frac{
8748 \lambda^{3} m^{4}\bigl(72 \lambda^{2} m^{2} + 27 \gamma^{2} M_{p}^{2}\bigr)
}{918330048 \lambda^{6}}.\notag
\end{align}

\bigskip

\noindent \textbf{2.} In regards to the expressions for $\mu_c$ and $G_{\rm eff}$ \cite{De_Felice_2016}
\begin{align*}
w_1 &= - 4 H G_4  - \psi^3 G_{3,X}, \\
w_2 &= w_1 + 4H [G_4-G_{4,X}\psi^2], \\
w_3 &= -2 \psi^2 
\\
w_4 &= 
       - 3 H^2 \left( 2 G_4 - 2 \psi^2 G_{4,X}  \right) \\
     &\quad - \frac{3}{2} H \psi^3 (G_{3,X} ),   \\
w_5 &= w_4 - \frac{3}{2} H (w_1 + w_2), \\
w_6 &= \psi 4 H G_{4,X}  -\psi^2 G_{3,X} \\
w_7 &= - 4 G_{4,X} \dot{H}  - G_{3,X} \dot{\psi}.
\end{align*}

\begin{align*}
    &\mu_1= \frac{\psi^2}{H}[(\dot w_1-2\dot w_2+Hw_1-3G_4H^2)w_3-2w_2(w_2+Hw_3)],\\
    &\mu_2=\psi[w_2^2+Hw_2w_3+\dot w_2w_3] +w_2[w_6\psi^2-\dot\psi w_3],\\
    &\mu_3=\frac{2\phi\mu_2}{Hw_3},\\
    &\mu_4=-\frac{1}{w_3}[\psi^3(w_6^2+2w_3w_7)+\psi^2(2w_2w_6+Hw_3w_6+w_3\dot w_6)],\\
    & \qquad \psi(w_2^2+Hw_2w_3+w_3(\dot w_2)-\dot \psi w_6)-2\dot \psi w_2w_3]\\
    &\mu_5=(w_1-2w_2)[\psi(w_1-2w_2)w_3\mu_4-2\psi w_2 \mu_2 ]\\
    &\qquad Hw_2[2w_2(\mu_1+w_3\mu_3)-w_1w_3\mu_3],\\
    &G_{\rm eff}=\frac{H(\mu_2\mu_3-\mu_1\mu_4)}{4\pi\psi\mu_5}.
\end{align*}

\bibliography{apssamp}
\end{document}